    \documentclass[aps,pre,showpacs,twocolumn,amsmath,floatfix]{revtex4}
   \usepackage{graphicx}
\newcommand\beq{\begin{equation}}
\newcommand\eeq{\end{equation}}
\newcommand\beqa{\begin{eqnarray}}
\newcommand\eeqa{\end{eqnarray}}

\newcommand\gh{\widetilde{\gamma}}
\newcommand\ph{\widetilde{p}}
\newcommand\Deltah{\widetilde{\Delta}}
\newcommand\etah{\widetilde{\eta}}

\begin{document}

\title{{Demixing can occur in binary hard-sphere mixtures with
negative non-additivity}}
\author{A. Santos}
 \email[]{andres@unex.es}
\homepage[]{http://www.unex.es/eweb/fisteor/andres/}
\author{M. L\'opez de Haro}
\email[]{malopez@servidor.unam.mx}
\homepage[]{http://miquiztli.cie.unam.mx/xml/tc/ft/mlh/}
\thanks{On sabbatical leave from Centro de Investigaci\'{o}n en Energ\'{\i}a, U.N.A.M., Temixco, Morelos 62580
(M{e}xico)} \affiliation{Departamento de F\'{\i}sica, Universidad de
Extremadura, E-06071 Badajoz, Spain}
%\thanks{}


\begin{abstract}
 {A binary fluid mixture of
non-additive hard spheres characterized by a size ratio
$\gamma=\sigma_2/\sigma_1<1$ and a non-additivity parameter
$\Delta=2\sigma_{12}/(\sigma_1+\sigma_2)-1$ is considered in
infinitely many dimensions. From the equation of state in the second
virial approximation (which is exact in the limit $d\to\infty$) a
demixing transition with a critical consolute point at a packing
fraction scaling as $\eta\sim d 2^{-d}$ is found, even for slightly
negative non-additivity, if
$\Delta>-\frac{1}{8}\left(\ln\gamma\right)^2$. Arguments concerning
the stability of the demixing with respect to freezing are
provided.}
\end{abstract}

\date{\today}
\pacs{64.75.+g,  61.20.Gy, 64.70.Ja, 82.60.Lf}
\maketitle

The crystallization of a hard-sphere fluid, first observed in
computer simulation \cite{WJ57} and at that time controversial, is a
clear and presently well established example of an entropy-driven
phase transition. Yet, up until now it has defied a rigorous
statistical mechanical proof. Other phase transitions governed by
entropy, in particular the phase separation in binary hard-core
mixtures, has only been proven \cite{FL92} for a simple
two-dimensional lattice model of two types of particles. In the
absence of an exactly solvable model in three dimensions, which
could provide insight into the detailed mechanisms leading to phase
separation in athermal systems such as hard-core mixtures, different
strategies have been adopted to address this problem. For instance,
to consider a geometry that leads to a non trivial volume-driven
phase separation, as in the case of a mixture of parallel hard cubes
 \cite{DF94}. Or else, to explore the situation in higher
spatial dimensions \cite{YSH00}.

Due to the interest (both theoretical and in practical applications)
of phase separation, the demixing problem in \textit{additive}
hard-sphere mixtures has received a lot of attention in the
literature. An analysis of the solution of the Percus--Yevick
equation for binary additive hard-sphere mixtures \cite{LR64} led to
the conclusion that no phase separation into two fluid phases
existed in these systems. The same conclusion is reached if one
considers the most popular equation of state proposed for such
mixtures, namely the
Boubl\'{\i}k--Mansoori--Carnahan--Starling--Leland (BMCSL)
\cite{BMCSL} equation of state. For a long time the belief was that
this was a true physical feature. Nevertheless, this belief started
to be seriously questioned after Biben and Hansen
 \cite{BibHan91} obtained fluid-fluid segregation in additive
hard-sphere mixtures out of the solution of the Ornstein--Zernike
equation with the Rogers--Young closure and subsequent work has
concentrated on attempting to clarify the issue. Coussaert and Baus
\cite{Coussaert} have proposed an equation of state with improved
virial behavior for a binary additive hard-sphere mixture that
predicts a fluid-fluid transition at very high pressures (metastable
with respect to a fluid-solid one). On the other hand, Regnaut
\textit{et al}.\ \cite{RDA01} have examined the connection between
empirical expressions for the contact values of the pair
distribution functions and the existence of fluid-fluid separation
in mixtures of additive hard spheres. Further, in the case of highly
asymmetric binary additive hard-sphere mixtures, the depletion
effect has been invoked as the physical mechanism behind demixing
(see for instance Ref.\ \onlinecite{DRE99} and the bibliography
therein). Finally, demixing in mixtures of additive hard spheres has
been examined recently \cite{LdHT04} using the low density expansion
of the pressure by adding successively one more exact virial
coefficient (up to the sixth virial coefficient). In this latter
work it was found that already within the second virial coefficient
approximation the fluid separates into two phases of different
composition with a lower consolute critical point.

In contrast to the above results, which have the drawback of having
been derived under various approximations and are therefore open to
question and controversy, non-additive hard-core systems with {\em
positive} non-additivity are certainly known to exhibit fluid-fluid
demixing,  although again this has not been rigorously proved in
general. The celebrated Widom--Rowlinson model \cite{WR70}
represents a prototype system that allows the detailed study of such
a phase transition, an aspect that continues to be of interest in
the recent literature \cite{JGMC97}. Provided fluid-fluid
segregation really occurs in additive hard-sphere mixtures, where
size asymmetry would be the source of the transition, it is not
unreasonable to expect that, given a certain degree of (high) size
asymmetry, demixing may also be present in the case of hard-sphere
mixtures with small {\em negative} non-additivity.  {This feature,
however, seems to have hardly received any attention \cite{REL01}.}
The purpose of this { Communication} is to address this problem and
provide { evidence in favor of the statement}  posed in the title of
the paper. To do so, we will not work in three-dimensional space,
but rather consider the limit of infinitely many dimensions in which
our result will be exact.

Although there had been a few earlier papers \cite{FI81,MT84}
dealing with hard spheres in dimensions greater than three, it was
after the pioneer work of Frisch \textit{et al}.\ \cite{FRW85} in
which they showed that the classical hard-sphere fluid in infinitely
many dimensions was amenable to full analytical solution, that
studies of high dimensional hard-sphere systems became common over
the years \cite{YSH00,L86,CB86,LM90,CFP91,FSL02,RLdeHS04}. The fact
that features such as the freezing transition are present in all
dimensionalities (except for $d=1$) and the parallel between high
spatial dimensions and limiting high density situations that seems
to exist in fluids (with the added bonus of greater mathematical
simplicity as one increases the number of dimensions) suggest that
one can gain insight into the thermodynamic behavior of say
three-dimensional systems by looking at a similar problem in higher
dimensions. As a matter of fact, the very elegant work of Carmesin
\textit{et al}.\ \cite{CFP91} has exploited this approach to
illustrate phase separation in a hard-sphere mixture with positive
non-additivity in infinite spatial dimensionality. We now consider
the more general case also for $d\to\infty$ in which the
non-additivity may take negative values.

Let us consider a binary mixture of non-additive hard spheres of
diameters $\sigma_1$ and $\sigma_2$  in $d$ dimensions. The hard
core of the interaction between a sphere of species $1$ and a sphere
of species $2$ is $\sigma_{12}\equiv
\frac{1}{2}(\sigma_1+\sigma_2)(1+\Delta)$, where the parameter
$\Delta$ characterizes the degree of non-additivity of the
interactions. Further assume (something that will become exact in
the limit $d\to\infty$ \cite{CFP91}) that the equation of state of
the mixture is described by the second virial coefficient only,
namely
\beq
p=\rho
k_BT \left[1+B_2(x_1)\rho\right],
\label{1}
\eeq
where $p$ is the
pressure, $\rho$ is the number density, $k_B$ is Boltzmann's
constant, $T$ is the temperature, and
\beq B_2(x_1)=v_d
2^{d-1}\left(x_1^2\sigma_{1}^d+x_2^2\sigma_{2}^d+2x_1x_2\sigma_{12}^d\right)
\label{2}
\eeq
is the second virial coefficient, $x_1$ and $x_2=1-x_1$ being the
mole fractions and  $v_d=(\pi/4)^{d/2}/\Gamma(1+d/2)$ being the
volume of a $d$-dimensional sphere of unit diameter. The Gibbs free
energy per particle is (in units of $k_BT$)
\beq
g=x_1\ln\left(x_1\rho\Lambda_1^d\right)+x_2\ln\left(x_2\rho\Lambda_2^d\right)
+2B_2(x_1)\rho,
\label{4}
\eeq
where $\Lambda_i$ ($i=1,2$) are the thermal de Broglie wavelengths.
{ Given a size ratio $\gamma \equiv \sigma_2/\sigma_1<1$, a value of
$\Delta$, and a dimensionality $d$, the consolute critical point
$(x_{1c},p_c)$ is the solution to $\left({\partial^2 g}/{\partial
x_1^2}\right)_p=\left({\partial^3 g}/{\partial x_1^3}\right)_p=0$,
provided of course it exists}. Then, from Eq.\ (\ref{1}) one can get
the critical density $\rho_c$.

We now introduce the scaled quantities
\beq
\ph\equiv 2^{d-1}v_d d^{-2}p\sigma_1^d/k_BT,\quad y\equiv
d^{-1}B_2\rho.
\label{new1}
\eeq
Consequently, Eqs.\ (\ref{1}) and (\ref{4}) can be rewritten
as
\beq
\ph=y\left(y+d^{-1}\right)/{\widetilde{B}_2},
\label{new2}
\eeq
\beq
g=\sum_{i=1}^2 x_i\ln\left(x_i\lambda_i\right)+ \ln \left({A_d
y}/{\widetilde{B}_2}\right)+2dy,
\label{new3}
\eeq
where $\widetilde{B}_2\equiv B_2/2^{d-1}v_d\sigma_1^d$,
$\lambda_i\equiv (\Lambda_i/\sigma_1)^d$, and $A_d\equiv
d/2^{d-1}v_d$. Next we take the limit $d\to\infty$ and assume that
the volume ratio $\gh\equiv \gamma^d$ is kept fixed and that there
is a (slight) non-additivity $\Delta= d^{-2}\Deltah$ such that the
scaled non-additivity parameter $\Deltah$ is also kept fixed in this
limit. Thus, the second virial coefficient can be approximated by $
\widetilde{B}_2=\widetilde{B}_2^{(0)}+\widetilde{B}_2^{(1)}d^{-1}+O(d^{-2})
$, where $\widetilde{B}_2^{(0)}=\left(x_1+x_2\gh^{1/2}\right)^2$ and
$\widetilde{B}_2^{(1)}=x_1x_2\gh^{1/2}K$, with $K\equiv
\frac{1}{4}\left(\ln\gh\right)^2+2 \Deltah$. Let us remark that { in
order to find  a consolute critical point}, it is essential to keep
the term of order $d^{-1}$ if $\Deltah \leq 0$. The equation of
state (\ref{new2}) can then be inverted to yield $
y=y^{(0)}+y^{(1)}d^{-1}+O(d^{-2}) $, with \mbox{$y^{(0)}=\sqrt{\ph
\widetilde{B}_2^{(0)}}$ and $y^{(1)}= -\frac{1}{2}
\left(1-y^{(0)}\widetilde{B}_2^{(1)}/\widetilde{B}_2^{(0)}\right)$}.
In turn, the Gibbs free energy (\ref{new3}) becomes $
g=g^{(0)}d+g^{(1)}+O(d^{-1}) $, with $g^{(0)}=2 y^{(0)}$,
$g^{(1)}=\sum_{i} x_i\ln\left(x_i\lambda_i\right)+\ln\left(A_d
y^{(0)}/\widetilde{B}_2^{(0)}\right)+2 y^{(1)}$, while the chemical
potentials $\mu_1=g+x_2\left(\partial g/\partial x_1\right)_p$ and
$\mu_2=g-x_1\left(\partial g/\partial x_1\right)_p$ are given by $
\mu_i=\mu_i^{(0)}d+\mu_i^{(1)}+O(d^{-1}) $, where
$\mu_1^{(0)}=2\ph^{1/2}$, $\mu_1^{(1)}=\ln\left(A_d x_1
\lambda_1\sqrt{\ph/\widetilde{B}_2^{(0)}}\right)-1/\sqrt{\widetilde{B}_2^{(0)}}+(x_2/x_1)
(\gh\ph)^{1/2}\widetilde{B}_2^{(1)}/\widetilde{B}_2^{(0)}$ and
$\mu_2$ is obtained from $\mu_1$ by the changes $x_1\leftrightarrow
x_2$, $\lambda_1\to \lambda_2/\gh$, $\gh\to 1/\gh$, $\ph\to \ph\gh$,
$\widetilde{B}_2\to \widetilde{B}_2/\gh$.

{ The coordinates of the critical point are readily found to be}
\beq
x_{1c}=\frac{\gh^{3/4}}{1+\gh^{3/4}},\quad
\ph_c=\frac{\left(1+\gh^{1/4}\right)^4}{4\gh K^2}.
\label{8}
\eeq
Note that $x_{1c}$ is independent of $\Deltah$. The coexistence
curve, which has to be obtained numerically, follows from the
conditions $\mu_i^{(1)}(x_A,\ph)=\mu_i^{(1)}(x_B,\ph)$ ($i=1,2$)
where $x_1=x_A$ and $x_1=x_B$ are the mole fractions of the
coexisting phases. Once the critical consolute point has been
identified in the pressure/concentration plane, we can obtain the
critical density. The dominant behavior of $\widetilde{B}_2$ at the
critical point is
$\widetilde{B}_2^{(0)}(x_{1c})=\gh/\left(1-\gh^{1/4}+\gh^{1/2}\right)^2$,
while
$y_c^{(0)}=\left(1+\gh^{1/4}\right)^2/2\left(1-\gh^{1/4}+\gh^{1/2}\right)K$.
Hence, the critical density readily follows after substitution in
the scaling relation given in Eq.\ (\ref{new1}). For our purposes it
is also convenient to consider the
 packing fraction defined as $\eta=v_d\rho\sigma_1^d\left(x_1+x_2\gh\right)$
 and its scaled version $\etah\equiv d^{-1} 2^d\eta$ \cite{footnote}. At the
critical point, this latter takes the nice expression
\beq
\etah_c=
\frac{\left(\gh^{1/8}+\gh^{-1/8}\right)^2}{K}.
\label{new9}
\eeq

\begin{figure}
\includegraphics[width=.95 \columnwidth]{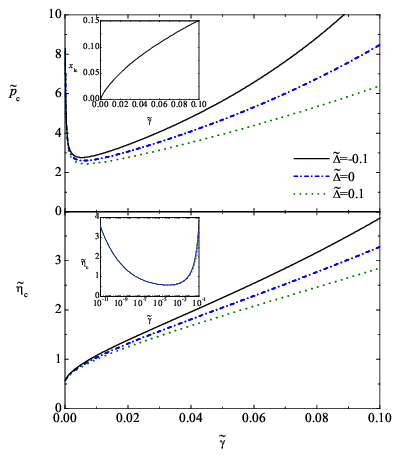}
\caption{(Color online) Plot of $x_{1c}$ (inset, upper panel),
$\ph_c$ (upper panel), and $\etah_c$ (lower panel) as functions of
$\gh$. The (scaled) critical  pressure and packing fraction are
displayed for three different values of the (scaled) non-additivity
parameter: $\Deltah=-0.1$ (solid lines), $\Deltah=0$ (dash-dotted
lines), and $\Deltah=0.1$ (dotted lines). Note the non-monotonic
dependence of $\ph_c$ and $\etah_c$ on $\gh$ (for the latter case,
see inset in the lower panel).
\label{fig1}}
\end{figure}
\begin{figure}
\includegraphics[width=.95 \columnwidth]{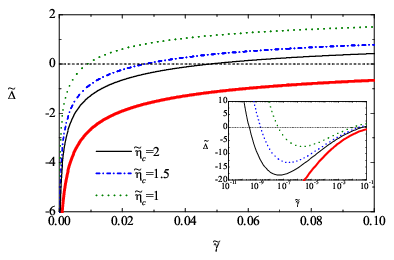}
\caption{(Color online) Plot of $\Deltah$ vs $\gh$ for different
values of the critical packing fraction: $\etah_c=2$ (thin solid
line), $\etah_c=1.5$ (dash-dotted line), and $\etah_c=1$ (dotted
line). The lowest thick solid line corresponds to the threshold
condition $\Deltah=-\frac{1}{8}\left(\ln\gh\right)^2$. A demixing
transition with the scaling properties of the text is only possible
for {mixtures with non-additivities and size ratios represented by}
points above the thick curve. On the other hand, a transition with a
(scaled) critical packing fraction smaller than a given value of
$\etah_c$ is only possible for points above the corresponding curve.
The threshold curve goes to minus infinity as $\gh \to 0$ but the
other curves have a minimum that depends on the choice of $\etah_c$
(see inset). {This implies that for \textit{extremely} asymmetric
mixtures ($\gh \to 0$) demixing at a given finite value of $\etah_c$
becomes possible only if the non-additivity is positive and
sufficiently large.}
\label{Delta}}
\end{figure}
Figure \ref{fig1} shows $x_{1c}$, $\ph_c$, and $\etah_c$ as
functions of $\gh$ and in the two latter cases for $\Deltah=-0.1$
{(negative non-additivity)}, $\Deltah=0$ (additive mixture) and
$\Deltah=0.1$ {(positive non-additivity)}. The previous results
clearly indicate that a demixing transition is possible {not only
for additive or positively non-additive mixtures but} even for
negative non-additivities. The only requirement is $K>0$, i.e.\
$\Deltah
>-\frac{1}{8}\left(\ln\gh\right)^2$ {or, equivalently, $\Delta
>-\frac{1}{8}\left(\ln\gamma\right)^2$}. The curve representing the
threshold situation $\Deltah=-\frac{1}{8}\left(\ln\gh\right)^2$ is
plotted in Fig.\ \ref{Delta}, where we have also displayed
$\Deltah$, as obtained from Eq.\ (\ref{new9}), as a function of
$\gh$ for three different values of the critical packing fraction:
$\etah_c=1$, $\etah_c=1.5$, and $\etah_c=2$. These choices for
$\etah_c$ are meant to be illustrative and have been taken after the
following considerations.

One may reasonably wonder whether the demixing we have obtained for
negative non-additivity will occur for
 packing fractions within the {stable} fluid regime and where the equation of
state is well represented by the second virial approximation. A
natural way to look into this issue would be to compare with the
close-packing value $\etah_{\text{cp}}$. Unfortunately,
$\etah_{\text{cp}}$ is not known in the case of mixtures.
Nevertheless, some insight about it can be gained by examining the
parallel case of a one-component fluid in infinitely many
dimensions. { In such a case, there exist known upper bounds for}
$\etah_{\text{cp}}$ \cite{FSL02}. Further, another (lower) estimate
can be given by taking the contributions of the second and third
virial coefficients to be of a similar order of magnitude. While
this is of course not conclusive,
 all these estimates {for $\etah_{\text{cp}}$} may be shown to diverge as $d\to\infty$, suggesting
that the fluid-fluid phase-separation may indeed take place.
Provided the (scaled) packing fraction at freezing $\etah_f$ is
different from zero, the demixing transition  may be stable and not
preempted by a fluid-solid transition. Again $\etah_f$ is unknown
but we may once more recur to the one-component case. For this
system, Colot and Baus  \cite{CB86} have conjectured that
$(\eta_f/\eta_{\text{cp}})^{1/d}$ becomes independent of $d$ for
high $d$. Further, from the analysis of the results in $d=3,4,5$
\cite{MT84,LM90}, and $d=7$  \cite{RLdeHS04} one finds that
$\etah_f\approx 1.3$. Since at freezing or melting the Helmholtz
free energies of the fluid and the solid should be of the same order
of magnitude, by considering the former given by the second virial
approximation and the latter as obtained from free volume theory
with the estimate $(\eta_f/\eta_{\text{cp}})^{1/d}\approx 0.8$, we
obtain the rough estimate $\etah_f\approx 2.3$. Irrespective of the
numbers, the point is that these results seem to confirm that
$\etah_f$ is different from zero and finite. Therefore, even if the
range of values of negative $\Deltah$ in which stable demixing
occurs is limited and restricted to highly asymmetric mixtures, as
indicated in Fig.\ \ref{Delta}, the important issue is that it is
certainly there. So the question of whether demixing can occur in
binary mixtures of hard spheres with negative non-additivity can be
given a positive answer.

{While the high dimensionality limit has allowed us to address the
problem in a mathematically simple and clear-cut way, the
possibility of demixing with negative non-additivity is not an
artifact of that limit. Demixing is known to occur  for positive
non-additive binary mixtures of hard spheres in three dimensions and
compelling evidence in the additive case exists, at least in the
metastable fluid region. Even though in a three-dimensional mixture
the equation of state is certainly more complicated than Eq.\
(\ref{1}) and the demixing transition reported here for negative
non-additivity is possibly metastable with respect to the freezing
transition, the main effects at work (namely the competition between
depletion due to size asymmetry and hetero-coordination due to
negative non-additivity) are also present. In fact, it is
interesting to point out that Roth  \textit{et al}.\ \cite{REL01},
using the approximation of an effective one-component fluid with
pair interactions to describe a binary mixture of non-additive hard
spheres and employing an empirical rule based on the effective
second virial coefficient, have also suggested that demixing is
possible for small negative non-additivity  and high size asymmetry.
Our exact results lend support to this suggestion based on
approximate calculations and confirm the fact that the limit
$d\to\infty$ allows one to get a caricature or toy model to
highlight features already present in real systems.}

%\acknowledgments
This work has been supported by Junta de
Extremadura  (Consejer\'{\i}a de Educaci\'on y Tecnolog\'{\i}a) and
Fondo Social Europeo under Project TEM04/0009. M.L.H. also wants to
thank D.G.A.P.A.-U.N.A.M. for a sabbatical grant. A.S. acknowledges
the financial support of Ministerio de Educaci\'on y Ciencia (Spain)
through Grant No. FIS2004-01399 (partially financed by FEDER funds).


\begin{thebibliography}{}
\bibitem{footnote}Note that, while the scaling implied in $\gh$ is the same as that of
Ref.\ \protect\cite{CFP91}, the one of the packing fraction is
different ($\eta_c\sim 2^{-d}d$ in our case in contrast to
$\eta_c\sim 2^{-d}$ in theirs) due to the different scaling of the
non-additivity parameter $\Delta$ ($\Delta\sim d^{-2}$ here and
$\Delta\sim d^{-1}$ in Ref.\ \protect\cite{CFP91}). As mentioned,
our scaling for $\Delta$ is necessary to deal with negative
non-additivity. But of course, for positive non-additivity, our
results are consistent with those of Ref.\ \protect\cite{CFP91}.
\end{thebibliography}

\end{document}